% Fusion of RSOS Models as a Coset Construction
% Michael Yu. Lashkevich
% hep-th/9605235
%%%%%%%%%%%%%%%%%%%%%%%%%%%%%%%%%%%%%%%%%%%%%%%%%%%%%%%%%%%%%%%%%
% This is preprint format MPPT.TEX  Version 1.00, 2.06.96       %
% by M.Lashkevich. No rights preserved.                         %
%%%%%%%%%%%%%%%%%%%%%%%%%%%%%%%%%%%%%%%%%%%%%%%%%%%%%%%%%%%%%%%%%
% This format is very similar to the Harvmac.tex but more flexible
% and spares all plain TeX commands, allows you to choose
% magnification, fonts etc. by yourself.
%%%%%%%%%%%%%%%%%%%%%%%%%%%%%%%%%%%%%%%%%%%%%%%%%%%%%%%%%%%%%%%%%
\catcode`@=11
\global\newcount\secno
\global\newcount\subsecno
\global\newcount\subsubsecno
\global\newcount\equationno
\global\newcount\refno
\global\newcount\@no
\secno=0\subsecno=0\subsubsecno=0\equationno=0\refno=0
\def\\{\cr}
\def\@draftleft#1{}
\def\@draftright#1{}
\overfullrule=0pt
\def\draft{\def\@draftleft##1{\leavevmode\vadjust{\smash{%
\raise3pt\llap{\eighttt\string##1~~}}}}%
\def\@draftright##1{\rlap{\eighttt~~\string##1}}%
\def\date##1{\leftline{\number\month/\number\day/\number\year\
\the\time}\bigskip}\overfullrule=5pt\normalbaselineskip=18pt
\normalbaselines}
\def\@the#1{\ifnum\the#1>0\relax\the#1\else\ifnum\the#1<0\relax
\@no=-\the#1\advance\@no'100\char\@no\fi\fi}
\def\@advance#1{\ifnum\the#1<0\global\advance#1 -1\relax
\else\global\advance#1 1\relax\fi}
\def\nsec#1\par{\bigskip\allowbreak\bigskip\@advance\secno
\subsecno=0\subsubsecno=0\equationno=0
\vbox{\secfont\noindent
\@the\secno. #1\medskip}\nobreak\noindent\ignorespaces}
\def\secadvance{\@advance\secno}
\def\sec#1#2\par{\bigskip\allowbreak\bigskip
\subsecno=0\subsubsecno=0\equationno=0
\if*#1\vbox{\secfont\noindent\ignorespaces#2\medskip}%
\else
\secno=#1
\vbox{\secfont\noindent\@the\secno. #2\medskip}\fi
\nobreak\noindent\ignorespaces}
\def\seclab#1{\xdef#1{\@the\secno}\@draftleft#1}
\def\nsubsec#1\par{\bigskip\@advance\subsecno
\subsubsecno=0\equationno=0
\vbox{\subsecfont\noindent\@the\secno.\@the\subsecno. #1\medskip}%
\nobreak\noindent\ignorespaces}
\def\subsecadvance{\@advance\subsecno}
\def\subsec#1#2\par{\bigskip
\subsubsecno=0\equationno=0
\if*#1\vbox{\subsecfont\noindent\ignorespaces#2\medskip}%
\else
\subsecno=#1
\vbox{\subsecfon\noindentt\@the\secno.\@the\subsecno. #2\medskip}\fi
\nobreak\noindent\ignorespaces}
\def\subseclab#1{\xdef#1{\@the\secno.\@the\subsecno}%
\@draftleft#1}
\def\nsubsubsec#1\par{\medskip\@advance\subsubsecno
\vbox{\subsubsecfont\noindent
\@the\secno.\@the\subsecno.\@the\subsubsecno. #1\medskip}%
\nobreak\noindent\ignorespaces}
\def\subsubsecadvance{\@advance\subsubsecno}
\def\subsubsec#1#2\par{\medskip
\if*#1\vbox{\subsubsecfont\noindent\ignorespaces#2\medskip}%
\else
\subsubsecno=#1
\vbox{\subsubsecfont\noindent
\@the\secno.\@the\subsecno.\@the\subsubsecno. #2\medskip}\fi
\nobreak\noindent\ignorespaces}
\def\subsubseclab#1{\xdef#1{\@the\secno.\@the\subsecno.\@the\subsubsecno}%
\@draftleft#1}
\def\eqlabel#1{\@advance\equationno
\ifnum\secno=0\xdef#1{\the\equationno}\else
\ifnum\subsecno=0\xdef#1{\@the\secno.\the\equationno}\else
\xdef#1{\@the\secno.\@the\subsecno.\the\equationno}\fi\fi
\eqno({\eqnofont #1})\@draftright#1}
\def\lnlabel#1{\global\advance\equationno1
\ifnum\secno=0\xdef#1{\the\equationno}\else
\ifnum\subsecno=0\xdef#1{\@the\secno.\@the\equationno}\else
\xdef#1{\@the\secno.\@the\subsecno.\the\equationno}\fi\fi
&({\eqnofont #1})\@draftright#1}
\def\eqadvance#1{\global\advance\equationno1
\ifnum\secno=0\xdef#1{\the\equationno}\else
\ifnum\subsecno=0\xdef#1{\@the\secno.\the\equationno}\else
\xdef#1{\@the\secno.\@the\subsecno.\the\equationno}\fi\fi}
\def\eqlabelno(#1#2){\eqno({\eqnofont #1#2})\@draftright#1}
\def\lnlabelno(#1#2){&({\eqnofont #1#2})\@draftright#1}
\newwrite\rfile
\def\nref#1#2{\global\advance\refno1\xdef#1{\the\refno}%
\immediate\write
\rfile{\noexpand\item{#1.}\noexpand\@draftleft\noexpand#1%
#2}}
\def\sref#1#2{\immediate\write
\rfile{\noexpand\item{#1.}\noexpand\@draftleft\noexpand#1%
#2}}
\def\refs#1#2 {\if.#2#2$^{\rm#1}$\spacefactor=\sfcode`.{}\space
\else\if,#2#2$^{\rm#1}$\spacefactor=\sfcode`,{}\space
\else\if;#2#2$^{\rm#1}$\spacefactor=\sfcode`;{}\space
\else\if:#2#2$^{\rm#1}$\spacefactor=\sfcode`:{}\space
\else\if?#2#2$^{\rm#1}$\spacefactor=\sfcode`?{}\space
\else\if!#2#2$^{\rm#1}$\spacefactor=\sfcode`!{}\space
\else
$^{\rm#1}$#2\space\fi\fi\fi\fi\fi\fi}
\def\Refs#1{$\rm#1$}
\def\reportno#1{\line{\hfil\vbox{\halign{\strut##\hfil\cr#1\crcr}}}}
\def\Title#1{\vskip3\bigskipamount\line{\titlefont
\hfil\vbox{\halign{\strut\hfil##\hfil\cr#1\crcr}}\hfil}%
\vskip2\bigskipamount}
\def\author#1{\centerline{\authorfont#1}\bigskip}
\def\address#1{\centerline{\vbox{\halign
{\strut\hfil\addressfont##\hfil\cr#1\crcr}}}%
\bigskip}
\def\abstract#1{{\narrower\abstractfont\null\bigskip\noindent\ignorespaces
#1\bigskip}}
\def\date#1{\leftline{#1}\bigskip}
\def\Tr{\mathop{\rm Tr}\nolimits}
\immediate\openout\rfile=refs.tmp
\font\seventeenrm=cmr17
\font\twelverm=cmr12  \font\eightrm=cmr8  \font\sixrm=cmr6
\font\seventeeni=cmmi10 scaled 1728
\font\twelvei=cmmi12  \font\eighti=cmmi8  \font\sixi=cmmi6
\font\seventeensy=cmsy10 scaled 1728
\font\twelvesy=cmsy10 scaled 1200 \font\eightsy=cmsy8 \font\sixsy=cmsy6
\font\seventeenbf=cmbx10 scaled 1728
\font\twelvebf=cmbx12 \font\eightbf=cmbx8 \font\sixbf=cmbx6
\font\seventeentt=cmbx10 scaled 1728
\font\twelvett=cmtt12 \font\eighttt=cmtt8
\font\seventeenit=cmti10 scaled 1728
\font\twelveit=cmti12 \font\eightit=cmti8
\font\seventeensl=cmsl10 scaled 1728
\font\twelvesl=cmsl12 \font\eightsl=cmsl8
\font\seventeenex=cmex10 scaled 1728
\font\twelveex=cmex10 scaled 1200
\def\tenpoint{\def\rm{\fam0\tenrm}%
\textfont0=\tenrm\scriptfont0=\sevenrm\scriptscriptfont0=\fiverm
\textfont1=\teni\scriptfont1=\seveni\scriptscriptfont1=\fivei
\textfont2=\tensy\scriptfont2=\sevensy\scriptscriptfont2=\fivesy
\textfont3=\tenex\scriptfont3=\tenex\scriptscriptfont3=\tenex
\textfont\itfam=\tenit\def\it{\fam\itfam\tenit}%
\textfont\slfam=\tensl\def\sl{\fam\slfam\tensl}%
\textfont\ttfam=\tentt\def\tt{\fam\ttfam\tentt}%
\textfont\bffam=\tenbf\scriptfont\bffam=\sevenbf
\scriptscriptfont\bffam=\fivebf\def\bf{\fam\bffam\tenbf}%
\def\small{\eightpoint}%
\def\large{\twelvepoint}%
\def\normalbaselines{\lineskip\normallineskip
  \baselineskip\normalbaselineskip \lineskiplimit\normallineskiplimit}
\setbox\strutbox=\hbox{\vrule height8.5pt depth3.5pt width0pt}%
\normalbaselines\rm}
\def\twelvepoint{\def\rm{\fam0\twelverm}%
\textfont0=\twelverm\scriptfont0=\eightrm\scriptscriptfont0=\sixrm
\textfont1=\twelvei\scriptfont1=\eighti\scriptscriptfont1=\sixi
\textfont2=\twelvesy\scriptfont2=\eightsy\scriptscriptfont2=\sixsy
\textfont3=\twelveex\scriptfont3=\twelveex\scriptscriptfont3=\twelveex
\textfont\itfam=\twelveit\def\it{\fam\itfam\twelveit}%
\textfont\slfam=\twelvesl\def\sl{\fam\slfam\twelvesl}%
\textfont\ttfam=\twelvett\def\tt{\fam\ttfam\twelvett}%
\textfont\bffam=\twelvebf\scriptfont\bffam=\eightbf
\scriptscriptfont\bffam=\sixbf\def\bf{\fam\bffam\twelvebf}%
\def\small{\tenpoint}%
\def\large{\seventeenpoint}%
\def\normalbaselines{\lineskip1.2\normallineskip
  \baselineskip1.2\normalbaselineskip \lineskiplimit1.2\normallineskiplimit}
\setbox\strutbox=\hbox{\vrule height10.2pt depth4.2pt width0pt}%
\normalbaselines\rm}
\def\seventeenpoint{\def\rm{\fam0\seventeenrm}%
\textfont0=\seventeenrm\scriptfont0=\twelverm\scriptscriptfont0=\eightrm
\textfont1=\seventeeni\scriptfont1=\twelvei\scriptscriptfont1=\eighti
\textfont2=\seventeensy\scriptfont2=\twelvesy\scriptscriptfont2=\eightsy
\textfont3=\seventeenex\scriptfont3=\seventeenex
\scriptscriptfont3=\seventeenex
\textfont\itfam=\seventeenit\def\it{\fam\itfam\seventeenit}%
\textfont\slfam=\seventeensl\def\sl{\fam\slfam\seventeensl}%
\textfont\ttfam=\seventeentt\def\tt{\fam\ttfam\seventeentt}%
\textfont\bffam=\seventeenbf\scriptfont\bffam=\twelvebf
\scriptscriptfont\bffam=\eightbf
\def\bf{\fam\bffam\seventeenbf}%
\def\small{\twelvepoint}%
\def\large{\seventeenpoint}%
\def\normalbaselines{\lineskip1.7\normallineskip
  \baselineskip1.7\normalbaselineskip \lineskiplimit1.7\normallineskiplimit}
\setbox\strutbox=\hbox{\vrule height14.5pt depth5.9pt width0pt}%
\normalbaselines\rm}
\def\eightpoint{\def\rm{\fam0\eightrm}%
\textfont0=\eightrm\scriptfont0=\sixrm\scriptscriptfont0=\fiverm
\textfont1=\eighti\scriptfont1=\sixi\scriptscriptfont1=\fivei
\textfont2=\eightsy\scriptfont2=\sixsy\scriptscriptfont2=\fivesy
\textfont3=\tenex\scriptfont3=\tenex\scriptscriptfont3=\tenex
\textfont\itfam=\eightit\def\it{\fam\itfam\eightit}%
\textfont\slfam=\eightsl\def\sl{\fam\slfam\eightsl}%
\textfont\ttfam=\eighttt\def\tt{\fam\ttfam\eighttt}%
\textfont\bffam=\eightbf\scriptfont\bffam=\sixbf
\scriptscriptfont\bffam=\fivebf\def\bf{\fam\bffam\eightbf}%
\def\small{\eightpoint}%
\def\large{\tenpoint}%
\def\normalbaselines{\lineskip.8\normallineskip
  \baselineskip.8\normalbaselineskip \lineskiplimit.8\normallineskiplimit}
\setbox\strutbox=\hbox{\vrule height7pt depth3pt width0pt}%
\normalbaselines\rm}
\def\small{\eightpoint}
\def\large{\twelvepoint}
\def\vfootnote#1{\insert\footins\bgroup
  \interlinepenalty\interfootnotelinepenalty
  \splittopskip\ht\strutbox % top baseline for broken footnotes
  \splitmaxdepth\dp\strutbox \floatingpenalty\@MM
  \leftskip\z@skip \rightskip\z@skip \spaceskip\z@skip \xspaceskip\z@skip
  \footnotefont\textindent{#1}\footstrut\futurelet\next\fo@t}
\def\titlefont{\seventeenpoint\bf}
\def\authorfont{\tenpoint\rm}
\def\addressfont{\tenpoint\it}
\def\abstractfont{\eightpoint\rm}
\def\secfont{\tenpoint\bf}
\def\subsecfont{\tenpoint\sl}
\def\subsubsecfont{\tenpoint\it}
\def\eqnofont{\tenpoint\rm}
\def\footnotefont{\eightpoint\rm}
\catcode`@=12
%%%%%%%%%%%%%%%%%%%%%%%%%%%%%%%%%%%%%%%%%%%%%%%%%%%%%%%%%%%%%%%%%
% This is the end of the macros MPPT.TEX
%%%%%%%%%%%%%%%%%%%%%%%%%%%%%%%%%%%%%%%%%%%%%%%%%%%%%%%%%%%%%%%%%
\hoffset=-1truecm
\def\W#1[#2,#3;#4,#5|#6]{W#1\biggl[\matrix{#5&#4\cr#2&#3}\Big|\,{#6}\,\biggr]}
\def\hatW{\expandafter\hat\W}
\def\T(#1,#2){\Theta_{#1}(#2)}
\def\(#1){(#1)_\infty}
\def\pointspace{\spacefactor=\sfcode`.{}\space}
\tolerance=1000
\reportno{LANDAU-96-TMP-2\\hep-th/9605235}
\Title{Fusion of RSOS Models as a Coset Construction}
\author{Michael Yu.~Lashkevich}
\address{Landau Institute for Theoretical Physics,\\
GSP-1, 117940 Moscow V-334, Russia}
\rightline{\sl Dedicated to the memory of Sasha Belov}
\abstract{
Using the vertex operator approach we show that fusion of the RSOS
models can be considered as a kind of coset construction which is very
similar to the coset construction of minimal models in conformal
field theory. We reproduce the excitation spectrum and $S$-matrix of
the fusion RSOS models in the regime III and show that their correlation
functions and form factors can be expressed in terms of those of the ordinary
(ABF) RSOS models.
}
\date{June 1996}
%
% References
%
\nref\dfjmn{B.~Davies, O.~Foda, M.~Jimbo, T.~Miwa, and A.~Nakayashiki,
	{\it Commun.\ Math.\ Phys.}\ {\bf 151}, 89 (1993)}
\nref\jmbook{M.~Jimbo and T.~Miwa, {\it
	Algebraic Analysis of Solvable Lattice
	Models}, American Mathematical Society, 1994}
\nref\frenresh{I.~B.~Frenkel and N.~Yu.~Reshetikhin, {\it Commun.\ Math.\
	Phys.}\ {\bf 146}, 1 (1992)}
\nref\etingof{P.~I.~Etingof, preprint {\tt hep-th/9312057} (December 1993)}
\nref\jmbos{M.~Jimbo, K.~Miki, T.~Miwa, and A.~Nakayashiki, {\it Phys.\
	Lett.}\ {\bf A168}, 256 (1992)}
\nref\luk{S.~Lukyanov, preprint RU-93-30,
	{\tt hep-th/9307196} (July 1993)}
\nref\lukpug{S.~Lukyanov and Ya.~Pugai, preprint
	RU-94-41, {\tt hep-th/9412128} (May 1994);
	preprint CLNS-96/1400, RIMS-1063,
	{\tt hep-th/9602074} (February, 1996)}
\nref\abf{G.~E.~Andrews, R.~J.~Baxter, and P.~J.~Forrester,
	{\it J.\ Stat.\ Phys.}\ {\bf 35}, 193 (1984)}
\nref\djkmo{E.~Date, M.~Jimbo, A.~Kuniba, T.~Miwa, and M.~Okado,
	{\it Nucl.\ Phys.}\ {\bf B290} [FS20], 231 (1987);
	{\it Adv.\ Stud.\ Pure Math.}\ {\bf 16}, 17 (1988)}
\nref\jmrsos{M.~Jimbo, T.~Miwa, and Y.~Ohta, {\it Int.\ J.\ Mod.\ Phys.}\
	{\bf A8}, 1457 (1993)}
\nref\crn{\v C.~Crnkovi\'c, G.~M.~Sotkov, and M.~Stanishkov,
	{\it Phys.\ Lett.}\ {\bf B226}, 297 (1989)}
\nref\lash{M.~Yu.~Lashkevich, {\it Mod.~Phys.~Lett.}\ {\bf A8}, 851 (1993)
	({\tt hep-th/9309071}); M.~Yu.~Lashkevich, {\it Int.\ J.\
	Mod.\ Phys.}\ {\bf A8}, 5673 (1993) ({\tt hep-th/9304116})}
\nref\bpz{A.~A.~Belavin, A.~M.~Polyakov, and A.~B.~Zamolodchikov,
	{\it Nucl.\ Phys.}\ {\bf B241}, 333 (1984)}
\nref\jmabf{O.~Foda, M.~Jimbo, T.~Miwa, K.~Miki, and A.~Nakayashiki,
	{\it J.\ Math.\ Phys.}\ {\bf 35}, 13 (1994)}
\nref\kadeishvili{A.~Kadeishvili, preprint LANDAU-96-TMP-1,
	{\tt hep-th/9604153} (April 1996)}
\nref\nakayashiki{A.~Nakayashiki, preprint {\tt hep-th/9409153}
	(August 1994)}
\nref\dotsenko{Vl.~S.~Dotsenko, {\it Adv. Stud. Pure Math.}\
	{\bf 16}, 123 (1988)}
\nref\lashbos{M.~Yu.~Lashkevich, {\it Int.\ J.\ Mod.\ Phys.}\
	{\bf A7}, 6623 (1992)}
\nref\bazhresh{V.~V.~Bazhanov and N.~Yu.~Reshetikhin, {\it Int.\ J.\
	Mod.\ Phys.}\ {\bf A4}, 115 (1989)}
\nref\douglass{M.~R.~Douglass, preprint CALT-68-1453 (1987)}
\nref\halpob{M.~B.~Halpern and N.~Obers, preprint LBI-32619,
	USB-PTH-92-24, BONN-HE-92-21, {\tt hep-th/9207071}
	(July 1992)}
\nref\lashcoset{M.~Yu.~Lashkevich, preprint LANDAU-92-TMP-1,
	{\tt hep-th/9301094} (October 1992)}

%%%%%%%%%%%%%%%%%%%%%%%%%%%%%%%%%%%%%%%%%%%%%%%%%%%%%%%%%%%%%%%%%%%%%%%%%%%
\nsec Introduction

The vertex operator (Kyoto) approach to solvable models of statistical
mechanics\refs{\dfjmn,\jmbook} makes it possible to express correlation
functions and form factors of local quantities in terms of correlation
functions of such highly nonlocal objects as vertex operators.
These new correlation
functions are solutions of the difference Knizhnik--Zamolodchikov
equation\refs{\frenresh,\etingof}.
Recently a great progress has been made by use of
the bosonic representation of vertex operators\refs{\jmbos-\lukpug}.
In particular, the bosonic representation for the ordinary (Andrews--Baxter--%
Forrester, ABF) RSOS models\refs{\abf} reveals the structure which is
very similar
to that of the conformal minimal models\refs{\lukpug}. This bosonic
representation is well investigated and allows one, in principle,
to find all correlators and form factors. On the other hand, the
$l$-fusion $(K+1)$-state RSOS model\refs{\djkmo} is known to be
described by the $(q,x)$-deformation ($q=x^{K+2}$)
of the coset construction\refs{\jmrsos}
$$
N_{kl}\sim{SU(2)_k\times SU(2)_l\over SU(2)_{k+l}},
\qquad k=K-l.
\eqlabel\klcoset
$$
It is known in conformal field theory that such coset models can
be also considered as coset constructions of minimal models%
\refs{\crn,\lash}
$$
{SU(2)_k\times SU(2)_l\over SU(2)_{k+l}}
\sim{M_kM_{k+1}\ldots M_{k+l-1}\over M_1\ldots M_{l-1}},
\eqlabel\mincoset
$$
where $M_n\sim N_{n1}$ is a minimal conformal model with the central
charge $c=1-6/(n+2)(n+3)$.
This representation allows one to express correlation functions in the
coset models  in terms of the correlation functions in the minimal
models.

In this letter a similar construction is proposed to express correlation
functions of the fusion RSOS models in terms of those of
the ordinary RSOS models.

%%%%%%%%%%%%%%%%%%%%%%%%%%%%%%%%%%%%%%%%%%%%%%%%%%%%%%%%%%%%%%%%%%%%%%%%%%%
\nsec Ordinary RSOS Models

The ordinary (ABF) RSOS models in the regime III
are identified
in the vertex operator approach with an elliptic deformation of the
holomorphic sector of the minimal models.

Recall that
the primary fields of the minimal model $M_k$ are given by the vertex
operators\refs{\bpz}
$$
\eqalign{
\span
\phi_{pq}^{(k)}(z)_m^{m'}{}_n^{n'}:
{\cal H}_{mn}\rightarrow{\cal H}_{m'n'},
\cr
\span
p,m,m'=1,2,\ldots,k+1;\quad q,n,n'=1,2,\ldots,k+2;
\cr
\span
|m-p|+1\leq m'\leq\min(m+p-1,k+3-m-p),
\cr
\span
|n-q|+1\leq n'\leq\min(n+q-1,k+4-n-q),
}\eqlabel\minvertices
$$
where ${\cal H}_{mn}$ is the state space generated by the Virasoro
algebra from the highest weight vector $\phi_{mn}(0)_1^m{}_1^n
|{\rm vac}\rangle$.

The fields $\phi_{12}(z)$ and $\phi_{21}(z)$
are of particular interest because they generate the whole set
of the vertex operators in fusion. In the deformed case
they have a simple meaning. The operator $\phi_{12}(z)_n^{n'}
\equiv\oplus_m\phi_{12}(z)_m^m{}_n^{n'}$ (type I vertex operator)
describes the half transfer matrix of the ordinary RSOS model%
\refs{\jmbook,\jmabf}. The operator $\phi_{21}(z)_m^{m'}\equiv
\oplus_n\phi_{21}(z)_m^{m'}{}_n^n$ (type II vertex operator)
represents the wave function of an elementary
excitation\refs{\jmbook,\lukpug}. Their commutation relations
are given by\refs{\lukpug}
$$
\eqalign{
\phi_{12}(z_1)_s^{n'}\phi_{12}(z_2)_n^s
&=\sum_{s'}\W^-_k[s,n;s',n'|z_1\over z_2]
\phi_{12}(z_2)_{s'}^{n'}\phi_{12}(z_1)_n^{s'},
\cr
\phi_{21}(z_1)_r^{m'}\phi_{21}(z_2)_m^r
&=\sum_{r'}\W^+_k[r,m;r',m'|z_1\over z_2]
\phi_{21}(z_2)_{r'}^{m'}\phi_{21}(z_1)_m^{r'},
\cr
\phi_{12}(z_1)_{m'}^{m'}{}_n^{n'}\phi_{21}(z_2)_m^{m'}{}_n^n
&=\tau\left(z_1\over z_2\right)
\phi_{21}(z_2)_m^{m'}{}_{n'}^{n'}\phi_{12}(z_1)_m^m{}_n^{n'}.
}\eqlabel\abfcomrel
$$
Here $W^-_k$ is the weight matrix of the RSOS model, $\tau(z)$ is the
elementary excitation spectrum, and $W^+_k$ is the $S$-matrix of two
elementary excitation. An $N$-particle excitation of the transfer
matrix $T^{(k)}(z)$ is described by a chain of integers $(m_0,m_1,m_2,\ldots,
m_{N-1},m_N)$, $m_{i+1}=m_i\pm1$, $m_i=1,2,\ldots,k+1$,
and a chain of complex numbers
(spectral parameters) $(z_1,z_2,\ldots,z_N)$, $|z_i|=1$.
It is given by an operator
$\phi_{21}(z_N)_{m_{N-1}}^1\ldots\phi_{21}(z_2)_{m_1}^{m_2}
\phi_{21}(z)_1^{m_1}$ in the sense of Refs.\ \Refs{\jmrsos,\jmbook}.
The respective eigenvalue of the transfer matrix
is $\tau(z/z_1)\tau(z/z_2)\ldots\tau(z/z_N)$. Quantities $m_0$ and
$m_N$ define boundary conditions and can be treated as topological
numbers of the solution. The boundary number $m_0=m$ ($m_N=m$)
means that we consider configurations stabilizing at the positive
(negative) infinity to the sequence \dots, $m$, $m+1$, $m$, $m+1$,
\dots

The coefficients $\tau$, $W^-_k$, and $W^+_k$ are expressed explicitly
in terms of the functions
$$
\eqalign{
\span
\T(p,z)=\(z;p)\(p/z;p)\(p;p),
\cr
\span
\(z;p_1,\ldots,p_N)=\prod_{n_1,\ldots,n_N=0}^\infty
\left(1-zp_1^{n_1}\ldots p_N^{n_N}\right)
}
$$
as follows\refs{\lukpug}:
$$
\eqalignno{
\span
\tau(z)=z^{-1/2}\T(x^4,xz)/\T(x^4,x^3z)=f(z^{-1})/f(z),
\lnlabel\taudef
\cr
\span
\eqalign{
\W^-_k[n_1,n_2;n_3,n_4|z]
&=z^{{1\over2}{k+2\over k+3}}{g^-_k(z^{-1})\over g^-_k(z)}
\hatW[n_1,n_2;n_3,n_4|z,{k+2\over k+3},x^{2(k+3)}],
\cr
\W^+_k[n_1,n_2;n_3,n_4|z]
&=z^{{1\over2}{k+3\over k+2}}{g^+_k(z^{-1})\over g^+_k(z)}
\hatW[n_1,n_2;n_3,n_4|z,{k+3\over k+2},x^{2(k+2)}],
}\lnlabel\abfw}
$$
where%
\footnote{$^a$}{We are working in the gauge with negative
$\W[n\mp1,n;n\pm1,n|z]$ in contrast to Ref.\ \Refs{\lukpug}.}
$$
\eqalign{
\hatW[n\pm1,n;n\pm1,n\pm2|z,a,q]
&=1,
\cr
\hatW[n\pm1,n;n\pm1,n|z,a,q]
&=z^{-a(1\pm n)}{\T(q,q^a)\T(q,q^{\mp an}z)\over\T(q,q^{\mp an})\T(q,q^az)},
\cr
\hatW[n\pm1,n;n\mp1,n|z,a,q]
&=q^{\mp a^2n}z^{-a}
{\T(q,q^{a(1\mp n)})\T(q,z)\over\T(q,q^{\mp an})\T(q,q^az)},
\cr
\hatW[n_1,n_2;n_3,n_4|z,a,q]
&=0\hbox{ otherwise},
}\eqlabel\hatws
$$
and
\def\xx{x^{2(k+2)},x^4}
\def\xy{x^{2(k+3)},x^4}
$$
\eqalign{
f(z)
&=z^{1/4}\(x^3z;x^4)/\(xz;x^4),
\cr
g^+_k(z)
&={\(z;\xx)\(x^{2(k+4)}z;\xx)\over\(x^2z;\xx)\(x^{2(k+3)}z;\xx)},
\cr
g^-_k(z)
&={\(x^2z;\xy)\(x^{2(k+4)}z;\xy)\over\(x^4z;\xy)\(x^{2(k+3)}z;\xy)}.
}
$$
The pairs $(n_1,n_2)$, $(n_2,n_3)$, $(n_3,n_4)$, and $(n_4,n_1)$
must satisfy  RSOS admissibility conditions:
$$
\eqalign{
(m,n)\hbox{ is a $\kappa$-admissible pair}
&\Leftrightarrow\cases{1\leq m,n\leq \kappa+1,\cr m=n\pm1,}
\cr
(m,n)\hbox{ is a ($\kappa,\lambda$)-admissible pair}
&\Leftrightarrow\cases{m+n=\lambda+2,\lambda+4,\ldots,2\kappa-\lambda+2,
\cr m-n=-\lambda,-\lambda+2,\ldots,\lambda.}
}\eqlabel\kadmiss
$$
Adjacent lattice variables are $(k+1)$-admissible, and
adjacent eigenstate labels are $k$-admissible.

Except usual relations as Yang--Baxter equation, crossing symmetry,
and unitarity, the matrices $W^\pm_k$ satisfy the relation crucial
in what follows:
$$
\sum_n\W^+_{k+1}[n_1,n_2;n,n_4|z]\W^-_k[n,n_2;n_3,n_4|z]=-\delta_{n_1n_3}
\eqlabel\wwp
$$
if $(n_1,n_2)$, $(n_1,n_4)$, $(n_2,n_3)$, and $(n_3,n_4)$ pairs are
$(k+1)$-admissible, and
$$
\eqalign{
\span
r^\pm_{n_1}\W^\pm_k[n_1,n_2;n_3,n_4|z]=
r^\pm_{n_3}\W^\pm_k[n_3,n_2;n_1,n_4|z],
\cr
\span
r^{\pm}_n=q_\mp^{n(a_\pm n-1)/2}
\T(q_\pm,q_\mp^n),
}\eqlabel\wwt
$$
where $a_\pm$, $q_\pm$ are
defined for $W^\pm_k$ as it follows
from Eq. (\abfw):
$$
\eqalign{
a_+&=(k+3)/(k+2),
\cr
a_-&=(k+2)/(k+3),
}\quad\eqalign{
q_+&=x^{2(k+2)}.
\cr
q_-&=x^{2(k+3)},
}
$$

Now let us define the fields $\phi_{p,p+1}(z)_m^{m'}{}_n^{n'}$ in the
$(q,x)$-deformed case. Note that there is no natural way to define
deformed fields $\phi_{pq}$, contrary to the situation in the conformal
field theory. We shall define the fields as follows%
\footnote{$^b$}{These vertex operators differ from those
proposed recently by Kadeishvili\refs{\kadeishvili}.
In the trigonometric limit $k\to\infty$
the operators $\phi_{p,p+1}(z)$
give the restricted versions of Nakayashiki's vertex operators
for the spin-inhomogeneous XXZ model\refs{\nakayashiki}. }
$$
\eqalign{
\phi_{pq}(z)_m^{m'}{}_n^{n'}
=\lim_{u_i\to z\atop v_j\to z}\sum {\cal N}_p(u,v)
&\phi_{21}(x^{2-p}u_1)^{m'}\phi_{21}(x^{4-p}u_2)
\ldots\phi_{21}(x^{p-2}u_{p-1})_m
\cr
\times
&\phi_{12}(x^{q-2}v_1)^{n'}\phi_{12}(x^{q-4}v_2)
\ldots\phi_{12}(x^{2-q}v_p)_n,
}\eqlabel\phipp
$$
where the normalization factor is given by
$$
{\cal N}_p(u,v)=\prod_{i<j}^{p-1}(g^+(x^{2(j-i)}u_j/u_i))^{-1}
\prod_{i<j}^{q-1}(g^-(x^{2(i-j)}v_j/v_i))^{-1}
\prod_i^{p-1}\prod_j^{q-1}f^{-1}(x^{p+q-2(i+j)}v_j/u_i).
\eqlabel\nfactor
$$
The summation is taken over all intermediate states allowed by
Eq. (\minvertices).

Note that all singular terms in this expression
cancel out. Indeed, according to Ref.\ \Refs{\lukpug} the singularities
appear in the pair products as follows
$$
\eqalign{
(g^+(x^2z/z'))^{-1}\phi_{21}(z')^m_{m\pm1}\phi_{21}(x^2z)^{m\pm1}_m
&=\pm{{\rm const}\over1-z/z'}+O(1),
\cr
(g^-(x^{-2}z/z'))^{-1}\phi_{12}(x^2z')^n_{n\pm1}\phi_{12}(z)^{n\pm1}_n
&=\pm{{\rm const}'\over1-z/z'}+O(1)
}
$$
as $z'\to z$. The summation over intermediate
states $m\pm1$, $n\pm1$ cancels the singular terms.

In the commutation relations of the operators $\phi_{pq}(z)$
$$
\phi_{p_1q_1}(z_1)_r^{m'}{}_s^{n'}\phi_{p_2q_2}(z_2)_m^r{}_n^s
=\sum_{r's'}\W^k_{p_1q_1p_2q_2}[rs,mn;r's',m'n'|z_1\over z_2]
\phi_{p_2q_2}(z_2)_{r'}^{m'}{}_{s'}^{n'}\phi_{p_1q_1}(z_1)_m^{r'}{}_n^{s'}
\eqlabel\phippcomrel
$$
the matrix $W^k_{p_1q_1p_2q_2}$ factors into three parts:
$$
\eqalign{
\span
\W^k_{p_1q_1p_2q_2}[rs,mn;r's',m'n'|z]=
\W^+_{k,p_1-1,p_2-1}[r,m;r',m'|z]
\W^-_{k,q_1-1,q_2-1}[s,n;s',n'|z]
\tau_{p_1q_1p_2q_2}(z),
\cr
\span
\W^\pm_{kll'}[n_1,n_2;n_3,n_4|z]=
\sum_{\{\kappa_{ij},\ i,j\neq0\}}\prod_{i=1}^l\prod_{j=1}^{l'}
\W^\pm_k[\kappa_{ij-1},\kappa_{i-1j-1};\kappa_{i-1j},\kappa_{ij}|
x^{\pm2(i-j)}z],
\cr
\span
\kappa_{0l}=n_1,\quad\kappa_{00}=n_2,\quad
\kappa_{l0}=n_3,\quad\kappa_{ll}=n_4,
\cr
\span
\tau_{p_1q_1p_2q_2}(z)=\prod_{i=1}^{p_1-1}\prod_{j=1}^{q_2-1}
\tau(x^{p_1+q_2-2(i+j)}z)
\prod_{i=1}^{p_2-1}\prod_{j=1}^{q_1-1}
\tau(x^{p_2+q_1-2(i+j)}z)
}\eqlabel\fusionWll
$$
The matrices $W^\pm_{kll'}$ are the result of $l\times l'$
fusion\refs{\djkmo} of the matrices $W^\pm_k$. The matrix $W^+_{kll'}$
is the weight matrix of the fusion RSOS model.
In the case $p_1=p_2$, $q_1=q_2$ the matrices $W^\pm_{kl}$ of the
fusion model with square faces appear:
$$
\W^\pm_{kl}[n_1,n_2;n_3,n_4|z]=\W^\pm_{kll}[n_1,n_2;n_3,n_4|z].
\eqlabel\fusionW
$$
We need to make a remark. Usually
the definition of the fusion $W$-matrices (\fusionWll) implies that
the summation is taken over all $\{\kappa_{ij}\}$ allowed by
unrestricted SOS admissibility conditions\refs{\dfjmn}. We suppose that
the summation is taken over $\{\kappa_{ij}\}$ satisfying
{\it the RSOS admissibility conditions}. But it can be checked
in simple cases (the first nontrivial one appears at $l=4$) that
the configurations with internal indices that do not
satisfy the RSOS admissibility conditions cancel each other.
The mechanism of this cancellation is very similar to that
of cancellation of inadmissible intermediate states in
the bosonic representation of minimal conformal models\refs{\dotsenko}.
In the general case we can argue in the following way. We know
that the RSOS vertex operators are represented in the boson
representation by the same operators as the SOS vertex operators do.
Then the commutation relations of vertex operators must be
given by the fusion weights in the usual sense.
But if the in-state is admissible, i.~e.\ belongs to a space
${\cal H}_{mn}$ with $m<k+1$, $n<k+2$, then the out-state and
all internal states in the Eq. (\phipp) are also admissible.
Then commutation relation must contain $W$ matrices with
RSOS admissibility condition in all internal indices.
Therefore, two types of $W$ matrices coincide.

%%%%%%%%%%%%%%%%%%%%%%%%%%%%%%%%%%%%%%%%%%%%%%%%%%%%%%%%%%%%%%%%%%%%%%%%%%%
\nsec Fusion RSOS Models: Type I Vertex Operators

Recall the structure of the coset models (\klcoset). The vertex
operators $\phi_{pp'q}(z)$ of the coset $N_{kl}$ are labelled by
three integers $p$, $p'$, and $q$ (see, for example, Ref.\ \Refs{\lashbos})
$$
\eqalign{
\span
\phi_{pp'q}^{(k,l)}(z)_m^{m'}{}_\mu^{\mu'}{}_n^{n'}:
{\cal H}_{m\mu n}\rightarrow{\cal H}_{m'\mu'n'},
\cr
\span
p=1,\ldots,k+1,\quad p'=1,\ldots,l+1,\quad q=1,\ldots,k+l+1,
\cr
\span
p+p'-q-1\in2{\bf Z}.
}\eqlabel\klvertex
$$
Here ${\cal H}_{m\mu n}$ is a decomposable but not irreducible
Virasoro algebra representation. The fusion rules are given by
$$
\eqalign{
|m-p|+1
&\leq m'\leq\min(m+p-1,k+3-m-p),
\cr
|\mu-p'|+1
&\leq\mu'\leq\min(\mu+p'-1,l+3-\mu-p'),
\cr
|n-q|+1
&\leq n'\leq\min(n+q-1,k+l+3-n-q).
}\eqlabel\klfusion
$$
Consider the field $\phi_{1,l+1,l+1}(z)_m^{m'}{}_\mu^{\mu'}{}_n^{n'}$
(or, equivalently, $\phi_{k+1,1,k+1}$). We will omit the indices
$m$, $m'$, $\mu$, and $\mu'$, because $m'=m$ and $\mu'=l+2-\mu$, and
they produce no nontrivial braiding. This field has a simple
representative in the coset construction of minimal models\refs{\lash}
$$
\phi_{1,l+1,l+1}\sim\phi_{12}^{(k)}\phi_{23}^{(k+1)}\ldots
\phi_{l,l+1}^{(k+l-1)}.
$$
More precisely, let in the deformed case
$$
\eqalign{
\phi(z)_n^{n'}
&\equiv\phi_{1,l+1,l+1}(z)_n^{n'}
\cr
&\sim\bigoplus_{\{m_i,m'_i\}}
r^{(k+1)}_{m_1}r^{(k+2)}_{m_2}\ldots r^{(k+l-1)}_{m_{l-1}}
\phi^{(k)}_{12}(z)_{m_0m_1}^{m'_0m'_1}
\phi^{(k+1)}_{23}(z)_{m_1m_2}^{m'_1m'_2}\ldots
\phi^{(k+l-1)}_{l,l+1}(z)_{m_{l-1}n}^{m'_{l-1}n'}
}\eqlabel\voI
$$
(the terms with $m'_0=m_0$ only are nonvanishing), $r_n^{(\kappa)}=r_n^+$
for $q_+=x^{2(\kappa+2)}$.
Using Eqs. (\wwp) and (\wwt) it is easy to check the following
commutation relation
$$
\phi(z_1)_s^{n'}\phi(z_2)_n^s=
(-)^{l(l-1)/2}
\sum_{s'}\W^-_{K-1,l}[s,n;s',n'|z_1\over z_2]
\phi(z_2)_{s'}^{n'}\phi(z_1)_n^{s'}.
\eqlabel\voIcomrel
$$
The functions $\tau(z)$ drop out from this expression.
The sign factor $(-1)^{l(l-1)/2}$ seems to be strange, but it has a clear
physical meaning. Namely, it is easy to check that
$$
\W^-_{Kl}[s,n;s',n'|1]=(-)^{l(l-1)/2}\delta_{ss'},
$$
if the pairs $(s,n)$, $(n,s')$, $(s',n')$, and $(n',s)$ are
$(K,l)$-admissible. It means that we need the factor $(-1)^{l(l-1)/2}$
to avoid the unphysical situation $\phi(z)^{n'}_s\phi(z)^s_n=0$.
Note that this argument excludes the simplest field
$\phi^{(k+l-1)}_{1,k+l+1}(z)$ from the candidates for the type I
vertex operators.

%%%%%%%%%%%%%%%%%%%%%%%%%%%%%%%%%%%%%%%%%%%%%%%%%%%%%%%%%%%%%%%%%%%%%%%%%%%
\nsec Fusion RSOS Models: Type II Vertex Operators

Let us try to find the type II vertex operators. Two main properties of
the type II vertex operators are: a) they commute with the type I vertex
operators up to a function; b) they are `elementary' fields, i.~e.\
they generate in fusion all operators commuting with the type I vertex
operators.
We conjecture that they coincide with fields $\phi_{221}(z)_m^{m'}
{}_\mu^{\mu'}{}_n^{n'}$. We will omit the indices $n$ and $n'$, because
$n'=n$. More precisely, we write
$$
\eqalign{
\span
\psi(z)_m^{m'}{}_\mu^{\mu'}\equiv
\phi_{221}(z)_m^{m'}{}_\mu^{\mu'},
\cr
\bigoplus_{\smash{\{n_i,n'_i\}}}r^{(2)}_{n_1}\ldots r^{(l-1)}_{n_{l-2}}
\xi(l,l-1)_{n_{l-1}}^{n'_{l-1}}r^{(l)}_{l+2-n_{l-1}}
\phi^{(1)}_{12}(z)_{1n_1}^{1n'_1}\phi^{(2)}_{23}(z)\ldots
\phi^{(l-1)}_{l-1,l}(z)_{n_{l-2}n_{l-1}}^{n'_{l-2}n'_{l-1}}
&
\cr
\times\phi_{221}(z)_{l+2-n_{l-1},\mu}^{l+2-n'_{l-1},\mu'}
&\sim\phi^{(k)}_{21}(z)_\mu^{\mu'}.
}\eqlabel\voII
$$
Here
$$
\eqalign{
\xi(\kappa,\lambda)_n^{n'}
&=\left[n+n'+\lambda\over2\right]_{\lambda+1}
\Big/\left[n-n'+2\kappa-\lambda+2\over2\right]_{\kappa+1-\lambda},
\cr
[u]
&=x^{(2u-\kappa-2)^2/4(\kappa+2)}
\Theta_{x^{2(\kappa+2)}}(x^{2u}),\qquad
[u]_t=\prod_{i=0}^{t-1}[u-i].
}
$$
Eq.\ (\voII) defines the operator $\psi(x)$ uniquely\refs{\lash}. Some
comments are necessary. The subscript $n_{l-1}$ in $\phi^{(l)}_{l-1,l}$
is paired with $l+2-n_{l-1}$ in $\phi_{221}$ in Eq.~(\voII), not with another
$n_{l-1}$. We need this to ensure the same commutation relation of both
sides of Eq.~(\voII) with the fermionic currents introduced below in
Sec.~5. To construct this pairing we have to use the Bazhanov--Reshetikhin
duality of the fusion $W$ matrices\refs{\bazhresh}. It relates the matrices
$W^-_{\kappa-1,\lambda,\lambda'}$ and
$W^-_{\kappa-1,\lambda,\kappa-\lambda'}$.
Applying this duality twice (once with respect
to $\lambda$ and once with respect to $\lambda'$) and assuming
$\lambda'=\lambda$ one can obtain
$$
\W^-_{\kappa-1,\lambda}[n_1,n_2;n_3,n_4|z]
=(-)^{\kappa+(\kappa+1)(\kappa-2\lambda)/2}
{\xi(\kappa,\lambda)_{n_3}^{n_4}\xi(\kappa,\lambda)_{n_2}^{n_3}
\over\xi(\kappa,\lambda)_{n_1}^{n_4}\xi(\kappa,\lambda)_{n_2}^{n_1}}
\W^-_{\kappa-1,\kappa-\lambda}[n_1,\kappa+2-n_2;n_3,\kappa+2-n_1|z].
$$
This gives the following commutation relations:
$$
\eqalign{
\psi(z_1)_r^{m'}{}_\rho^{\mu'}\psi(z_2)_m^r{}_\mu^\rho
&=-\sum_{r'\rho'}\W^+_{k}[r,m;r',m'|z_1\over z_2]
\W^+_{l}[\rho,\mu;\rho',\mu'|z_1\over z_2]
\psi(z_2)_{r'}^{m'}{}_{\rho'}^{\mu'}\psi(z_1)_m^{r'}{}_\mu^{\rho'},
\cr
\phi(z_1)_n^{n'}\psi(z_2)_m^{m'}{}_\mu^{\mu'}
&=\tau(z_1/z_2)\psi(z_2)_m^{m'}{}_\mu^{\mu'}\phi(z_1)_n^{n'}.
}\eqlabel\voIIcomrel
$$
>From these commutation relations the following picture of excited
states can be drawn. Any ($N$-particle) eigenvector of the
transfer matrix $T_{kl}(z)$ is labelled by a chain of pairs
of positive integers
$\{(m_0,\mu_0),(m_1,\mu_1),\ldots,(m_{N-1},\mu_{N-1}),
(m_N,\mu_N)\}$, $m_{i+1}=m_i\pm1$, $\mu_{i+1}=\mu_i\pm1$,
$m_i\leq k+1$, $\mu_i\leq l+1$,
and by a chain of complex numbers $\{z_1,\ldots,z_N\}$, $|z_i|=1$.
The eigenvector is represented by the operator
$\psi(z_N)_{m_{N-1}}^{m_N}{}_{\mu_{N-1}}^{\mu_N}\ldots
\psi(z_1)_{m_0}^{m_1}{}_{\mu_0}^{\mu_1}$. The associated
eigenvalue of the transfer matrix $T^{(k,l)}(z)$
is $\tau(z/z_N)\ldots\tau(z/z_1)$. The pairs $(m_0,\mu_0)$ and
$(m_N,\mu_N)$ define conditions at the infinity. A pair $(m,\mu)$
corresponds to the configuration \dots, $m+\mu-1$, $m+l+1-\mu$,
$m+\mu-1$, $m+l+1-\mu$, \dots\pointspace
This coincides with the results of Ref.\ \Refs{\jmrsos}.

\nsec Notes on Traces of Vertex Operators

It is known that conformal blocks of coset models are given
by solutions of some linear algebraic equations\refs{\douglass-\lashcoset}.
The traces of vertex operators satisfy similar equations.
Here we make some notes concerning these equations.

Consider for example the zero-point correlation function of vertex
operators $\chi^{(k,l)}_{pp'q}=\Tr_{{\cal H}_{pp'q}}(x^{4D_{k,l}})$
($D_{k,l}$ is the grading
operator with the same spectrum as the Virasoro generator $L_0$ in
conformal field theory). In terms of these quantities
the local state probabilities are expressed\refs{\djkmo}.
The local spin here is $q$,
whereas $p$ and $p'$ are defined by the boundary conditions.
It is evident that $\chi^{(k,l)}_{pp'q}$ satisfies the equation
$$
\sum_{p'\{\mu\}}
\chi^{(1)}_{\mu_0\mu_1}\ldots\chi^{(l-1)}_{\mu_{l-2},l+2-p'}
\chi^{(k,l)}_{pp'q}
=\sum_{\{m\}}
\chi^{(k)}_{pm_1}\chi^{(k+1)}_{m_1m_2}\ldots\chi^{(k+l-1)}_{m_{l-1}q},
$$
where $\chi^{(k)}_{pq}=\Tr_{{\cal H}_{pq}}(x^{4D_k})$ is the zero-point
correlation function of an ABF model. Evidently, these equations
do not define the amount $\chi^{(k,l)}_{pp'q}$ uniquely. We need more
equations. To find them consider the field in the denominator
of our coset construction
$$
\eqalign{
\span
\sigma_s(z)={1\over N_s}
\sum_{\lambda\lambda'}r^{(s)}_\lambda\phi^{(s)}_{12}(z)_\lambda^{\lambda'}
\phi^{(s+1)}_{21}(z)_\lambda^{\lambda'},
\cr
\span
\sigma_s(z_1)\sigma_s(z_2)=-\sigma_s(z_2)\sigma_s(z_1),
\qquad
\sigma_{s\pm1}(z_1)\sigma_s(z_2)=\tau(z_1/z_2)
\sigma_s(z_2)\sigma_{s\pm1}(z_1),
\cr
\span
\sigma_{s'}(z_1)\sigma_s(z_2)=\sigma_s(z_2)\sigma_{s'}(z_1),
\quad|s'-s|>1.
}\eqlabel\EEiniteq
$$
Here $N_s$ is a normalization factor. It can be chosen, for example,
so that $\langle\sigma_s(1)\sigma_s(0)\rangle=1$.
The fields $\sigma_s(z)$ can be considered as additional currents.
In conformal field theory it has the conformal dimension $1/2$. In the
conformal limit where the coset construction is studied rigorously it can be
checked that it coincides with the field $\sigma_{k+s}(z)$
in the numerator
$$
\sigma_s(z)=\sigma_{k+s}(z).
$$
We conjecture that this identity holds in the deformed case.
Let us introduce the trace
$$
\Sigma^{(s)}_{n_1n_2n_3}=
\Tr_{{\cal H}^{(s)}_{n_1n_2}\otimes{\cal H}^{(s+1)}_{n_2n_3}}
\left(x^{4(D_s+D_{s+1})}\sigma_s(x^{-2}z)\sigma_s(z)\right),
$$
which is independent of $z$. We obtain $l-1$ equations
$$
\sum_{p'\{\mu\}}
\chi^{(1)}_{\mu_0\mu_1}\ldots
\Sigma^{(s)}_{\mu_{s-1}\mu_s\mu_{s+1}}\ldots
\chi^{(l-1)}_{\mu_{l-2},l+2-p'}\chi^{(k,l)}_{pp'q}
=\sum_{\{m\}}
\chi^{(k)}_{pm_1}\chi^{(k+1)}_{m_1m_2}\ldots
\Sigma^{(k+s)}_{m_{s-1}m_sm_{s+1}}\ldots
\chi^{(k+l-1)}_{m_{l-1}q}.
\eqlabel\EEaddeq
$$
We have $l$ equations for $l+1$ variables $\chi^{(k,l)}_{pp'q}$.
But $\chi^{(k,l)}_{pp'q}=0$ unless $p+p'-q-1$ is odd. Hence, we have
only $\lfloor(l+1)/2\rfloor$ or $\lfloor(l+2)/2\rfloor$ variables.

The similar procedure is necessary for other trace calculations.
The origin of this feature is related to the fact that the set of
the Virasoro charges does not form the complete chiral algebra of the
product $M_1\ldots M_{l-1}$. We need additional currents. Fermionic
fields $\sigma_s(z)$ play the role of these currents. But they cannot
enter the character of this model in the usual way $\exp(i\theta
({\rm charge}))$, and we need instead to consider additional traces
$\Sigma^{(s)}_{n_1n_2n_3}$.

\nsec Conclusion

We considered the elliptic deformation of the coset construction (\klcoset)
and gave evidences that it describes the fusion RSOS models in the
framework of the Kyoto approach to integrable models of statistical
mechanics. Despite absence of a rigorous proof it gives a plain recipe
for calculation of correlation functions.

\sec * Acknowledgments

I am grateful to A.~Kadeishvili, T.~Miwa, and Ya.~Pugai for
fruitful discussions.
I discussed some initial ideas of this paper
with Sasha Belov in the last months of his life, and I dedicate it
to his memory.

This work was supported in part by RFBR under the grant 95-02-05985,
and by INTAS under the grants CNRS 1010-CT93-0023 and INTAS-93-1939.

\bigskip\allowbreak\bigskip\immediate\closeout\rfile
\vbox{\secfont\noindent References\bigskip}\nobreak
\catcode`@=11\input refs.tmp\catcode`@=12\bigskip
\end